 \documentstyle[12pt,epsfig,twoside,fleqn,espcrc1]{article}

\newcommand{\AmS}{{\protect\the\textfont2
  A\kern-.1667em\lower.5ex\hbox{M}\kern-.125emS}}

\title{Probing the EOS of dense neutron-rich matter with high-energy
radioactive beams} 
\author{Bao-An Li\footnote{email: Bali@astate.edu}\address{Department of Chemistry and Physics, \\
Arkansas State University, P.O. Box 419,\\
State University, AR 72467-0419, USA}}

\begin{document}

\maketitle

\begin{abstract}
Nuclear reactions induced by high energy radioactive beams 
create a transient state of nuclear matter with high density and 
appreciable neutron to proton asymmetry. This will provide a unique opportunity
to explore novel properties of dense neutron-rich matter and the isospin-dependence 
of the nuclear equation of state (EOS). Here we study the $\pi^-/\pi^+$ ratio as a
probe of the EOS of dense neutron-rich matter.
\end{abstract}

\section{INTRODUCTION}
With isospin symmetry conserving interactions, the ${\rm EOS}$ of 
neutron-rich matter of isospin asymmetry 
$\delta\equiv (\rho_n-\rho_p)/(\rho_n+\rho_p)$ can be written 
as (see e.g., \cite{book}), 
$e(\rho,\delta)= e(\rho,0)+E_{sym}(\rho)\delta^2+{\cal O}(\delta^4)$,
where $e(\rho,\delta)$ and $e(\rho,0)$ is the energy per nucleon in 
isospin asymmetric and symmetric nuclear matter, respectively. 
The density dependence of the symmetry energy $E_{sym}(\rho)$ is the 
most important issue about the EOS of neutron-rich matter\cite{pra97,bali98,lat00,bom01,sci}. 
In fact, to determine the $E_{sym}(\rho)$ has been a longstanding goal of 
extensive research with various microscopic and/or phenomenological 
models over the last few decades. However, the predicted $E_{sym}(\rho)$ are 
still rather divergent especially at high densities. Theoretical results 
can be roughly classified into two groups, i.e., a group where the 
$E_{sym}(\rho)$ rises monotonically with density and one in which it begins 
to fall above about $2\rho_0$\cite{brown,mar}. The high density behaviour of 
$E_{sym}(\rho)$ is actually considered by 
some astrophysicists as the most uncertain property of dense 
neutron-rich matter\cite{kut}. 

The importance of probing the EOS of neutron-rich matter 
lies in both nuclear physics and astrophysics. First of all,
new structures of dripline nuclei, such as the appearance of new magic numbers, 
and the reaction mechanisms of radioactive nuclei are all strongly 
influenced by the isospin-dependence of the nuclear EOS. 
It also has profound consequences on many important issues in 
astrophysics\cite{pra97,bali98,lat00,bom01,sci}. These include nucleosynthesis 
in pre-supernova evolution of massive stars, mechanisms of supernova 
explosions, neutrino flux, kaon condensations and the hadron-QGP phase transition in
neutron stars, etc. 
\begin{figure}[htp] 
\vspace{-2.cm}
\centering \epsfig{file=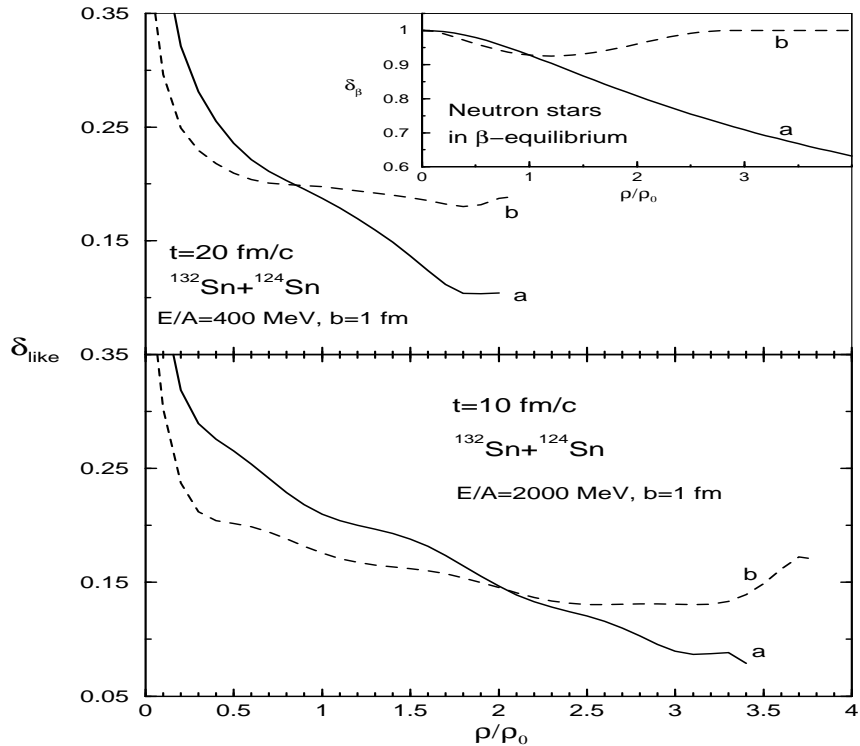,width=10cm,height=11cm,angle=-90} 
\vspace{-1cm}
\caption{The isospin asymmetry-density correlations in the central $^{132}Sn+^{124}Sn$ at 400 MeV/A (upper) 
and 2 GeV/A (lower) reaction with the
nuclear symmetry energy $E^a_{sym}$ and $E^b_{sym}$, respectively.} 
\label{fig8}
\end{figure}
\section{ISOSPIN ASYMMETRY IN NEUTRON STARS AND HEAVY-ION COLLISIONS}
In a neutron star, several of its main properties are affected significantly by the correlation
between the density and the proton fraction $x_{\beta}$ at $\beta$ equilibrium. 
The latter is determined by    
$
\hbar c(3\pi^2\rho x_{\beta})^{1/3}=4E_{\rm sym}(\rho)(1-2x_{\beta}).
$   
The equilibrium proton fraction is therefore entirely determined by 
the $E_{sym}(\rho)$. We use the following two forms of 
the symmetry energy for an illustration 
$
E^a_{sym}(\rho)\equiv E_{sym}(\rho_0)u $ and $
E^b_{sym}(\rho)\equiv 0.5E_{sym}(\rho_0)u\cdot(3-u),
$
where $u\equiv\rho/\rho_0$. 
The values of $\delta_{\beta}=1-2x_{\beta}$ corresponding to 
these two forms of symmetry energy are shown in the insert of Fig.\ 1.
With the $E^{b}_{sym}(\rho)$, 
the $\delta_{\beta}$ is $1$ for $\rho/\rho_0\geq 3$, 
indicating that the neutron star has become a pure neutron matter at 
these high densities. On the contrary, with the $E^{a}_{sym}(\rho)$, 
the neutron star becomes more proton-rich as the density increases.
To see the connection between neutron stars and heavy-ion collisions, 
using a nuclear transport model\cite{li96,li97,li00} we study in Fig.\ 1 
the isospin asymmetry $\delta_{like}$ in central $^{132}Sn+^{124}Sn$ reactions\cite{li02}. 
Effects due to the different symmetry energies are clearly 
revealed especially at high densities. An astonishing similarity is seen in the resultant 
$\delta-\rho$ correlations for neutron stars and heavy-ion collisions 
since the same underlying nuclear ${\rm EOS}$ is at work in both cases. 

\section{$\pi^-/\pi^+$ PROBE OF THE EOS of DENSE NEUTRON-RICH MATTER}
It is qualitatively easy to understand why the $\pi^-/\pi^+$ ratio in heavy-ion collisions 
can be used to extract crucial information about the {\rm EOS} of neutron-rich matter. 
On one hand, within the $\Delta$ resonance model for pion production from first-chance 
independent nucleon-nucleon collisions, the primordial $\pi^-/\pi^+$ ratio 
is $(5N^2+NZ)/(5Z^2+NZ)\approx (N/Z)^2$. It is thus sensitive to the isospin 
asymmetry $(N/Z)_{dense}$ of dense matter in the participant region of heavy-ion collisions. 
On the other hand, within the statistical model for pion production\cite{nature}, 
the $\pi^-/\pi^+$ ratio is proportional to 
${\rm exp}\left[(\mu_n-\mu_p)/T\right]$, 
where T is the temperature. The difference in nucleon chemical potentials 
$\mu_n-\mu_p$ is proportional to the strength of the symmetry potential\cite{li03}. Therefore from
both pictures the $\pi^-/\pi^+$ ratio is useful for extracting the EOS of dense 
neutron-rich matter.
\begin{figure}[htp] 
\vspace{-0.5 cm}
\centering \epsfig{file=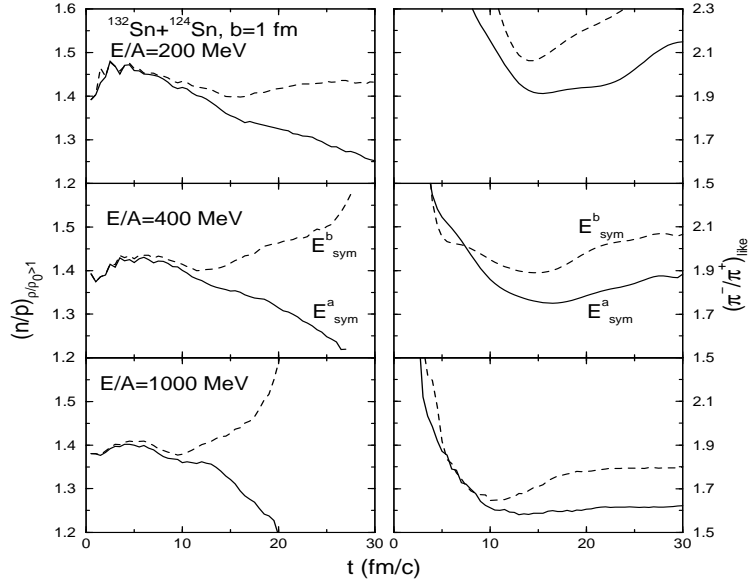,width=8cm,height=10cm,angle=-90} 
\vspace{-1cm}
\caption{The evolution of neutron/proton ratio of dense region (left) 
and $\pi^-/\pi^+$ ratio (right) in the central $^{132}Sn+^{124}Sn$ reactions 
at $E_{beam}/A=200, 400$ and 1000 MeV.} 
\label{fig2}
\end{figure}

Shown in Fig.\ 2 are the isospin asymmetry of the high density region 
$(n/p)_{\rho\geq \rho_0}$ (left) and the $(\pi^-/\pi^+)_{like}$ ratio (right)
$
(\pi^-/\pi^+)_{like}\equiv (\pi^-+\Delta^-+\frac{1}{3}\Delta^0)/
(\pi^++\Delta^{++}+\frac{1}{3}\Delta^+)
$ 
as a function of time for the central $^{132}Sn+^{124}Sn$ reaction at 
beam energies from 200 to 1000 MeV/nucleon. This ratio naturally 
becomes the final $\pi^-/\pi^+$ ratio at the freeze-out 
when the reaction time $t$ is much longer 
than the lifetime of the delta resonance $\tau_{\Delta}$.
The $(\pi^-/\pi^+)_{like}$ ratio is rather high in the early 
stage of the reaction because of the large numbers of neutron-neutron 
scatterings near the surfaces where the neutron skins of the colliding nuclei overlap. 
It is seen that a variation of about 
30\% in the $(n/p)_{\rho/\geq \rho_0}$ due to the different $E_{sym}(\rho)$ 
results in about 15\% change in the final $\pi^-/\pi^+$ ratio.
It has thus an appreciable response factor of 
about 0.5 to the variation of n/p ratio and is approximately 
beam energy independent. Therefore, the $(\pi^-/\pi^+)_{like}$ ratio 
is a sensitive probe of the high density behaviour of nuclear symmetry energy.
Shown in Fig.\ 3 is the reaction system dependence of the  $\pi^-/\pi^+$ ratio. 
The symmetry energy effects increase with the isospin asymmetry energy as one expected.  
\begin{figure}[htp]
\vspace{-1.2cm} 
\centering \epsfig{file=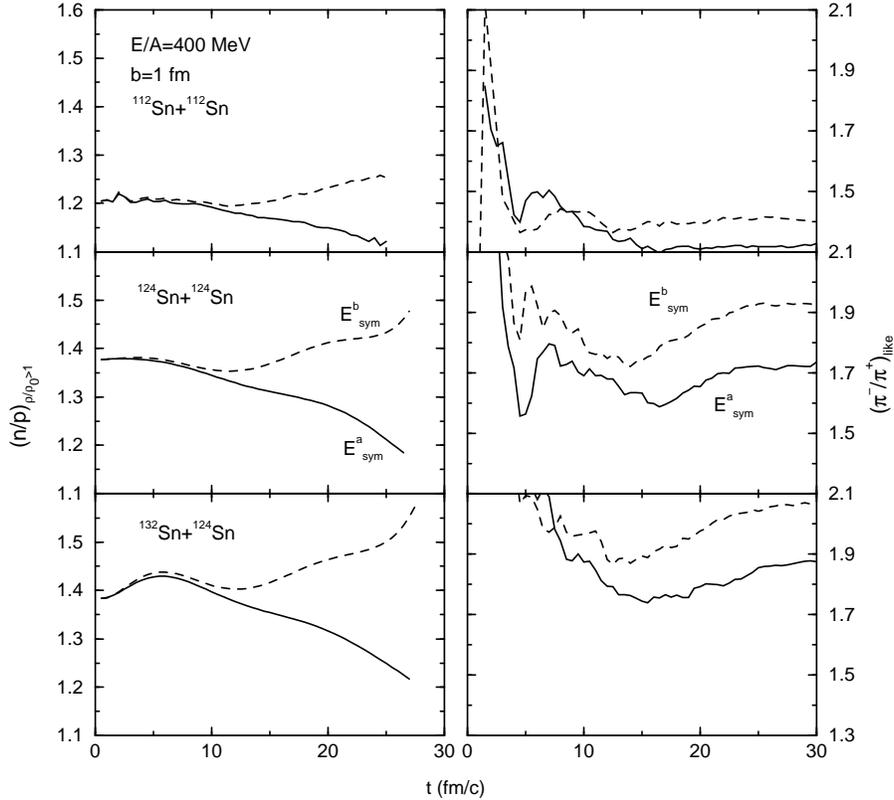,width=11.cm,height=12cm,angle=-90} 
\vspace{-1.cm}
\caption{Left panels: average neutron/proton ratio in the whole space with densities higher than the normal nuclear matter density as a function of time.
Right panels: the $(pi^-/\pi^+)_{like}$ ratio as a function of time for the three reactions.} 
\label{fig3}
\end{figure}

In summary, within an isospin-dependent transport model 
we found that the $\pi^-/\pi^+$ ratio is a promising
probe of the EOS of dense neutron-rich matter. 
This work was supported by the National Science Foundation under 
grant No. PHY-0088934 and PHY-0243571.

\end{document}